\input{aipcheck}

\documentclass[
    ,final            
  ]
  {aipproc}

\layoutstyle{8x11single}

\usepackage{amssymb}
\usepackage{graphics}
\usepackage{epsfig}

\begin{document}

\title{AGB STARS \& PRESOLAR GRAINS.}


\author{M. Busso}{
  address={INFN \& University of Perugia, Perugia, Italy}
}
\author{O. Trippella}{
  address={INFN \& University of Perugia, Perugia, Italy}
}
\author{E. Maiorca}{
  address={INAF - Arcetri Astrophysical Observatory, Firenze, Italy \& INFN - Section of Perugia, Perugia, Italy}
}
\author{S. Palmerini}{
  address={Departamento de F\`isica Te\`orica y del Cosmsos, Universidad de Granada, Granada, Spain}
}

\begin{abstract}
Among presolar materials recovered in meteorites, abundant SiC and Al$_{2}$O$_{3}$ grains of AGB origins were found. They showed records of C, N, O, $^{26}$Al and s-element isotopic ratios that proved invaluable in constraining the nucleosynthesis models for AGB stars \cite{zin,gal}. In particular, when these ratios are measured in SiC grains, they clearly reveal their prevalent origin in cool AGB circumstellar envelopes and provide information on both the local physics and the conditions at the nucleosynthesis site (the H- and He-burning layers deep inside the structure). Among the properties ascertained for the main part of the SiC data (the so-called {\it mainstream} ones), we mention a large range of $^{14}$N/$^{15}$N ratios, extending below
the solar value \cite{mar}, and $^{12}$C/$^{13}$C ratios $\gtrsim$ 30. Other classes of grains, instead, display low carbon isotopic ratios ($\gtrsim 10$) and a huge dispersion for N isotopes, with cases of large $^{15}$N excess. In the same grains, isotopes currently feeded by slow neutron captures reveal the characteristic pattern expected from this process at an efficiency slightly lower than necessary to explain the solar main s-process component. Complementary constraints can be found in oxide grains, especially Al$_{2}$O$_{3}$ crystals. Here, the oxygen isotopes and the content in $^{26}$Al are of a special importance for clarifying the partial mixing processes that are known to affect evolved low-mass stars. Successes in modeling the data, as well as problems in explaining some of the mentioned isotopic ratios through current nucleosynthesis models are briefly outlined.
\end{abstract}

\maketitle

\section{Introduction.}

Moderately massive stars ($0.8 \lesssim M/M_{\odot} \lesssim 8$), after the end of core H-burning, grow in luminosity while the surface temperature decreases and the convective envelope expands: for these reasons they are called Red Giants. In the classical HR diagram they populate a well defined sequence, called Red Giant Branch (RGB). Later, in the evolutionary stages that follow He exhaustion in the core, the expansion of the envelope resumes, making such stars extremely luminous and cool; the track they climb  in the diagram is called Asymptotic Giant Branch (and we speak of "AGB stars"). At the very end of their evolution, AGB stars are powered by two thermonuclear shells, burning H and He alternatively; He, in particular, burns in recurrent explosions called {\it thermal pulses}. After most of them, the (upward and downward) expansion of the envelope is re-established, bringing repeatedly to the surface the products of H and He burning in a global phenomenon called {\it third dredge up} (TDU). As carbon is the main product of He burning, AGB stars, with an initially O-rich composition, gradually become enriched in carbon (and in other He-burning products, including elements beyond iron, synthesized by slow neutron captures and called {\it s-elements}). For suitable values of the envelope mass (evolving in time due to efficient mass loss) AGB stars can finally achieve a surface abundance ratio (by number) C/O $\gtrsim 1$, in which case they become carbon stars of the most common class, called C(N). This is though to occur for initial masses in the limited range 1.7 $\lesssim M/M_{\odot} \lesssim$ 3, below which the star remains O-rich. Notice that mass loss rates are very uncertain and still the object of unsafe parameterizations. This is has become a crucial source of doubts on the details of AGB nucleosynthesis and for this reason several projects dedicated to observe the AGB winds on an homogeneous basis are underway (see e.g. \cite{gua,tos}).

Spectroscopy reveals that during the gradual changes in composition of AGB stars described above the spectral types evolve along the sequence \cite{wal}: M$\rightarrow$MS$\rightarrow$S$\rightarrow$C(N). The composition itself determines the type of condensates that form in the outer and cooler parts of the envelope. Thus, O-rich AGB stars form oxides (especially Al$_2$O$_3$) and silicates, while C(N)-stars are parents to silicon carbide (SiC) and graphite dust grains. Above a total mass of $M \gtrsim 3-4 M_{\odot}$ stars cannot become carbon rich, not only because the mass of the envelope to pollute is huge, but also because the hot temperature at its base (T $>8\times 10^7$ K) induces  CN-cycling, so that any carbon dredged-up can be burnt into nitrogen and/or oxygen. This process is known as {\it hot bottom burning, or HBB}. Observed s-element enrichments \cite{ab1} are in rather good agreement with stellar and nucleosynthesis models for AGB stars \cite{cris}. Also the appearance of other He-burning products like F can be accounted for \cite{ab2}, so that the AGB evolutionary stages are thought to be rather well understood. Spectroscopically, C(N) stars show $^{12}$C/$^{13}$C ratios that are typically $\gtrsim 30$, averaging at $\sim 60$, in a good agreement with most SiC grains (generally indicated as forming the ``mainstream'').

Other carbon-rich red giants exist, of more unclear origin. Among them, SC-type stars
show molecular bands indicating a C/O ratio very close to unity; the s-element enhancements are not always present and the $^{12}$C/$^{13}$C ratios range from about $3-4$ (close to the CNO-cycling equilibrium) up to $\sim 100$. A few of them are super Li-rich \cite{ab3}, with Li abundances exceeding the average values of C(N) giants by $4-5$ orders of magnitude. They also present large F enhancements \cite{ab2}: e.g. values of [F/Fe] $\sim 1$ (where [A/B] $=$ log$(N_A/N_B)-$log$(N_A/N_B)_{\odot}$) were observed in C-rich objects of solar metallicity ([Fe/H] $\sim 0$). Furthermore, there is recent observational evidence \cite{gua} that C(N) stars are on average less luminous than SC giants, which fact suggests progenitor stellar masses of $3-4$ M$_{\odot}$ for these last, against an average of 1.7$-$2 M$_{\odot}$ for C(N) ones. This sheds doubts on the presence of SC stars in the spectral sequence going from M to C(N) types. Another subgroup of carbon stars, called C(J), shows very strong features of $^{13}$C-bearing molecules, indicating $^{12}$C/$^{13}$C $ \lesssim 15$. They do not show s-element enhancements and most of them ($\sim 80\%$) have moderate Li enrichments \cite{ab4}. Their relation with the spectral sequence quoted above is also unclear. An important fraction ($\sim 30\%$) of C(J) stars shows emission lines in the infrared, usually associated with silicate dust. This might appear as contradictory, considering their C-rich composition, but chemical kinetics actually allows for this possibility in special cases \cite{codust}. These emissions seem to be originated from an O-rich disc in a binary system \cite{ram}, suggesting that at least some of these stars be indeed binaries.

A remarkable property of evolved red giants, especially those of low mass, remaining O-rich throughout their evolution, is the presence of partial mixing phenomena (often collectively called {\it deep mixing}) linking the envelope to active nuclear regions of the star. The most evident traces of this occurrence found in the grains are displayed by oxide and SiC compounds, through C, N and O isotopic admixtures showing enrichment in those nuclei (like $^{13}$C, $^{14}$N and $^{18}$O) that are more efficiently affected (in production or destruction) by proton captures. A similar evidence is displayed by excesses of $^{26}$Mg, which reveal the original presence in the stellar envelope of the unstable nucleus $^{26}$Al.  In all these cases the anomalous abundances of the grains inform us on the parameters characterizing the partial mixing events.

In the following two sections we present a limited number of examples of how the abundances measured in
either O- or C-rich presolar materials can be compared with nucleosynthesis models for Low Mass AGB stars, thus providing constraints to the models. Sometimes these constraints help us in  specifying the correct model
in terms of stellar mass, metallicity, efficiency of mixing etc. Sometimes however, the grains reveal details that the models cannot reproduce completely, thus opening a window for future research.

\section{Presolar Grains and AGB Processes I. Intermediate Elements}

\begin{figure}[t!!]
\includegraphics[width=0.8\textheight]{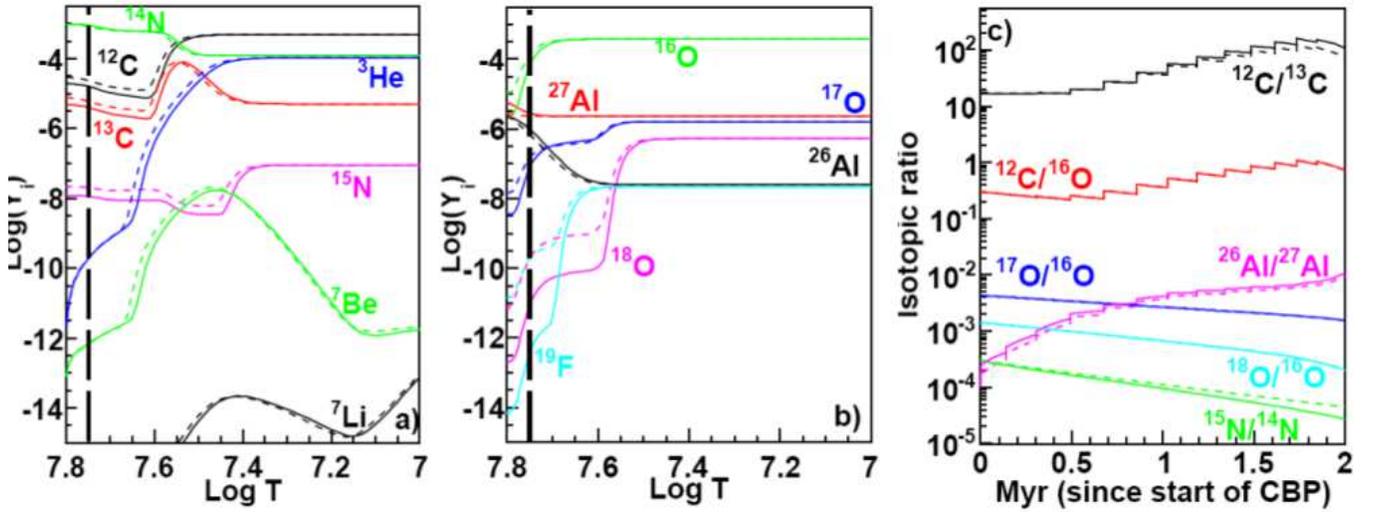}
\caption{Left and central panels: the chemical composition of the radiative region of a 2 $M_{\odot}$, solar metallicity AGB star, from the base of the H-shell to the base of the convective envelope. Vertical dashed lines indicate the layers of maximum energy production ($T=T_H$). Right panel: the predicted temporal evolution of isotopic ratios in the circumstellar envelope where dust grains form. It results from the occurrence of deep mixing (with $\log T_{\rm H} - \log T_P = 0.1$ and $\dot M = 10^{-6} M_{\odot}/yr$). The stepwise trend derives from the further occurrence of intermittent TDU episodes, bringing fresh $^{12}$C to the surface. The stellar model considered is the same as in the previous panels. The comparison between solid and dashed lines reveals the effects of adopting either recent reaction rate improvements (solid) \cite{ade} or the previous NACRE compilations (dashed). See \cite{pal} for details.}
\end{figure}

\begin{figure}[t!!]
\includegraphics[width=0.8\textheight]{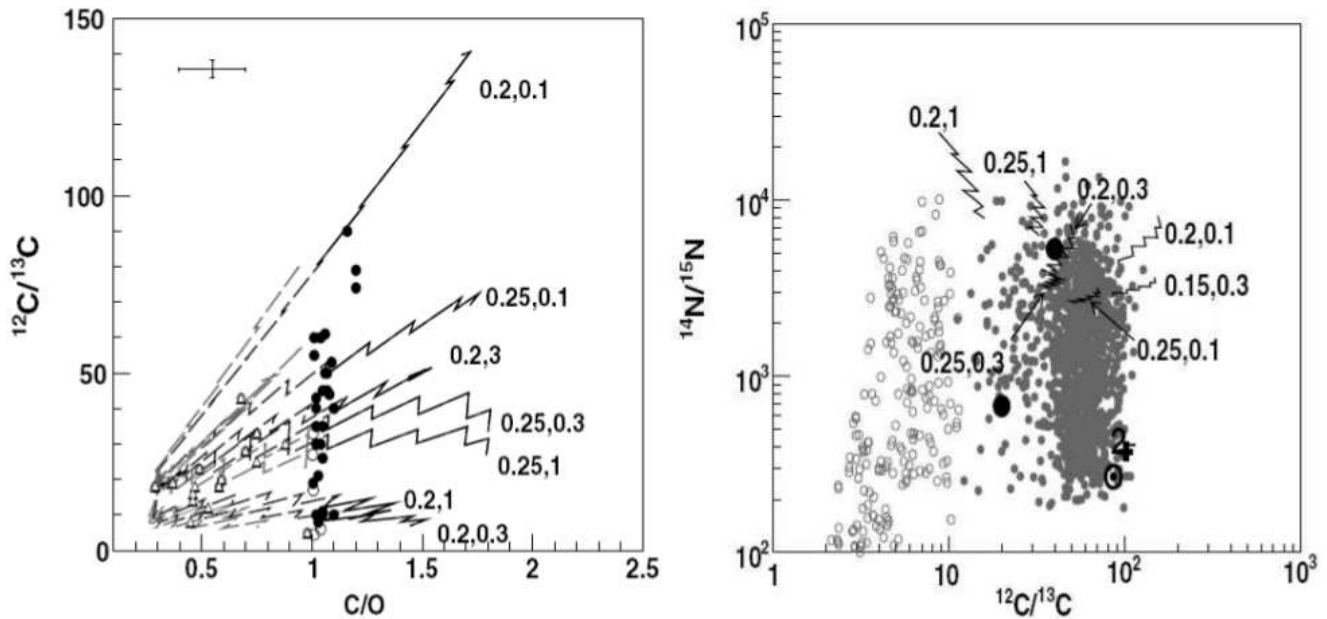}
\caption{Left panel: observations of C/O and $^{12}$C/$^{13}$C ratios in AGB stars of types MS, S (white triangles), SC (open circles), and C(N) (filled circles), as compared to model curves from stars of 1.5 $M_{\odot}$ (gray curves) and 2 $M_{\odot}$ (black curves) at solar metallicity, experiencing extramixing and TDU. Continuous lines refer to the C-rich phases, dashed lines to the O-rich ones. The labels indicate the choices for $\Delta = \log T_H - \log T_P$ and $\dot M$ in unit of $10^{-6} M_{\odot}$/yr. The area of the data is well covered by the models, indicating that the parameter choices should be rather typical of real AGB stars. Right panel: the $^{14}$N/$^{15}$N ratios of SiC grains recovered from pristine meteorites, as a  function of their $^{12}$C/$^{13}$C ratios. Open symbols represent the A+B grains, gray symbols the so-called mainstream ones (see the WUSTL Presolar Database: http://presolar.wustl.edu/~pgd/). Model curves are from a 2 $M_{\odot}$ star of solar metallicity; only the final C-rich phases are plotted. Again, the model parameters are indicated. The range of carbon isotopic ratios of mainstream grains and of some A+B grains is reproduced, but our $^{14}$N/$^{15}$N ratios are always much larger than solar. The full dots represent stellar measurements by in C-rich circumstellar envelopes and confirm that high values are typical of evolved stars. See \cite{pal,b10} for comments}
\end{figure}

\begin{figure}[t!]
\includegraphics[width=0.8\textheight]{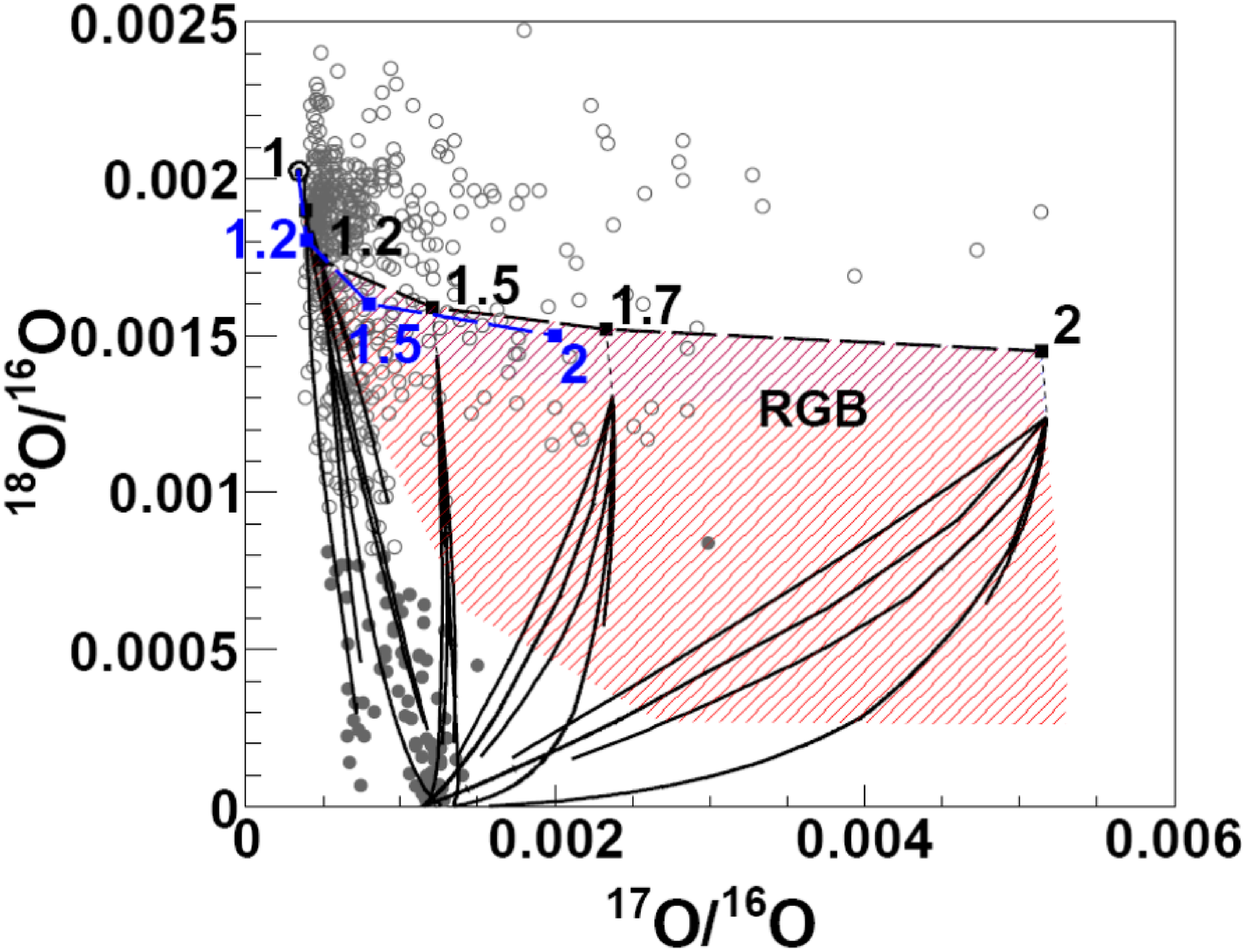}
\caption{The dashed black curve represents the oxygen isotopic mix in the envelope of solar-metallicity stars approaching the RGB phase. Full squares indicate the mass (from 1 to 2 $M_{\odot}$) of the considered models. As a comparison, the dashed blue and blue symbols show the composition for the masses 1.2, 1.5 and 2$M_{\odot}$ with NACRE rates. The shaded red area shows the range of values covered by a very efficient case of deep mixing, on the RGB ($\log T_P$ = 0.15 and $\dot M = 3\times10^{-7} M_{\odot}$/yr). Continuous lines show model results for extra-mixing on the AGB, different lines refer to different mixing parameters. The occurrence of a moderate extra-mixing episode during the RGB phase with $\log T_P = 0.2$ and $\dot M = 10^{-7} M_{\odot}$/yr is also considered (short-dashed grey lines). The grey data points refer to measurements in presolar grains (from the WUSTL Presolar Database, http://presolar.wustl.edu/pgd/). We plot those grains of group 1 (open circles) and of group 2 (filled circles) that are poor in $^{}$O, a sign of having experienced exposures to H-burning conditions. It is clear that deep mixing on the RGB, even with extreme efficiency, cannot reproduce the data of group 2 grains. Models for $M = 1.7 M_{\odot}$ explain essentially all the data for $^{18}$O-poor grains. In particular, group 2 grains mainly derive from very low-mass stars (below 1.5$M_{\odot}$).}
\end{figure}

As mentioned in the previous section, isotopic ratios of intermediate elements measured in presolar grains are of high relevance in fixing the parameters of the so called {\it deep mixing} episodes occurring in evolved stars. These partial mixing phenomena involve the exposure of a limited fraction of the envelope to high temperatures where nuclear processing occur; they are required to explain the evidence for CNO admixtures at the surface of evolved red giants not accounted for by traditional stellar models \cite{pal}. These phenomena are usually parameterized as a function of two free parameters, assuming that mixing  can be described either as a diffusive process or as a circulation (the results of the two approaches are often the same, as shown in \cite{nol}, except for the cases where the velocity of mixing is important). Assuming that we can describe the matter transport through a circulation, as an illustrative example, the parameters are the temperature $T_P$ of the deepest layers achieved by mixing and the circulation rate $\dot M$. In general, the first parameter is introduced through the logarithmic difference $\Delta = \log T_{\rm H} - \log T_{P}$, where $T_{\rm H}$ is the temperature at which the maximum energy is released by the H-burning shell.

Figure 1 (left and central panels) shows a typical trend of the abundances in the radiative layers above the H-burning shell and below the convective envelope of a low mass AGB star. By mixing these layers partially with the envelope and by accounting also for the TDU episodes mentioned above, one can obtain the trend shown in the right panel of the figure, where surface isotopic ratios are shown. (Here a 2 $M_{\odot}$ star of solar metallicity was used as an example). The choices for the two free parameters are indicated in the figure label. With techniques similar to the one described here, we can fit the composition of the photosphere and of the circumstellar envelope, hence also the one of solid particles condensating there.

The examples shown before and a direct comparison of theoretical predictions with stellar spectroscopy and with  C-rich grain measurements indicate how a considerable set of isotopic abundances in presolar grains can be well fitted by nucleosynthesis models for AGB stars, by suitably choosing the free parameters still affecting them.
This is in particular shown by Figure 2. In the left panel (see the label in the figure) we present data for the C/O and the $^{12}$C/$^{13}$C ratios in O-rich and C-rich AGB stars, as compared to model sequences of the envelope composition as a function of time. We can see how the spread in the abundance ratios shown by real stars can be well explained in terms of models for AGB envelopes affected by deep mixing processes at different efficiencies. The resulting spread in the carbon isotopic ratio at C/O = 1 (typical of C-stars) is from 10 to 100 and is the same measured for mainstream SiC grains.

In the right panel one also sees that most of the model sequences have their equivalent in zones of the plot showing the $^{12}$C/$^{13}$C and $^{14}$N/$^{15}$N ratios of mainstream grains. However, Figure 2 (right panel in particular), also reveals that there remain large unexplained areas covered by the grain data, which now call for an interpretation. This is so especially for the nitrogen isotopic admixtures measured. They extend to values of the $^{14}$N/$^{15}$N ratio so low that cannot be accomodated in our present scenario for normal star evolution. This is so even when one assumes nitrogen isotopic ratios at stellar birth different than solar \cite{nol}. Very recently, spectroscopic observations of C and N isotopes in C-stars by \cite{hed} added more surprise to the already complex scenario. These authors indeed showed that isotopic ratios of these crucial elements at odds with stellar nucleosynthesis predictions, but similar to those observed in several presolar SiC grains, are present directly in the envelopes of some peculiar C-rich giants of types SC and C(J). This emphasizes that presolar grains do come mostly from red giants and may sometimes indicate the correct stellar composition even when our models completely fail at explaining them. The puzzle is now studied looking for possible anomalous binary-star paths, to form common-envelope objects and then {\it apparently single} C-stars with abundances very different from what normal single-star evolution foresees; but this interpretation is still very far from being quantitative.

Another remarkable example of how the grains can teach us the way in which stars really behave is displayed in Figure 3. It refers to oxygen isotopic ratios of Al$_2$O$_3$ grains, especially those which are very poor in $^{18}$O, usually identified as being ``of group 2'' \cite{nit}. As this nucleus is very fragile against proton captures, low   $^{18}$O abundances indicate materials exposed to efficient H-burning, hence the presence of deep mixing linking the envelope to ashes of the H-burning shell (see Figure 1). The dashed line labeled from 1 to 2 (intended in units of solar masses) shows the typical composition of stars in this mass range at the beginning of the RGB branch. As is clear the grains are much poorer than that in $^{18}$O and group around the ratio $^{17}$O/$^{16}$O $\simeq$ 0.001, typical of H-burning equilibrium at high temperature. Both these facts suggest extensive mixing with H-rich material and as the model curves show, specific models of deep mixing, with suitable choices for the parameters, succeed in explaining the data. The mixing required is however much more effective than provided by diffusion mechanisms, like the popular {\it thermohaline mixing} \cite{egg1}. Hence the grains here also indicate that a faster process is required (see \cite{pal} for details).

\section{Presolar Grains and AGB Processes II. The s-Process}

\begin{figure}[t!]
\includegraphics[width=0.8\textheight]{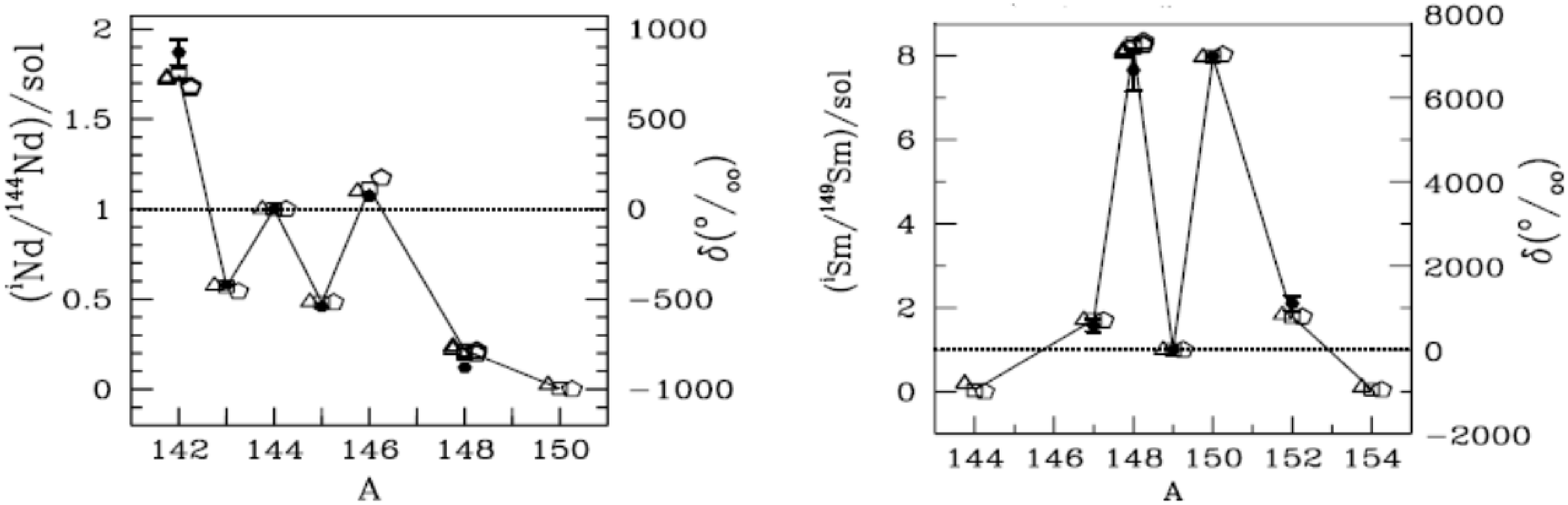}
\caption{Left panel: Nd isotopic ratios normalized to solar in the He-shell material cumulatively dredged up into the envelope, adopting $^{144}$Nd as reference nucleus. The lines derive from neutron-capture nucleosynthesis models in AGB stars of solar metallicity, with different efficiencies for the neutron release (different open symbols). Only cases with C/O > 1 in the envelope are shown as the measurements refer to SiC grains. The measurements themselves are from \cite{ric}. Right panel: Sm isotopic ratios normalized to solar in the He-shell material cumulatively dredged up into the envelope, adopting $^{149}$Sm as reference isotope. The models and the source for the measured data as the same as in the previous panel.}
\end{figure}

Several works in the nineties addressed the isotopic ratios of trace elements heavier than
iron in SiC grains. Generally, they belong to the so-called {\it ``s''} (or ``slow'') process.
This is a sequence of neutron captures, starting from iron isotopes and reaching up to Pb and Bi,
which proceed along the valley of $\beta$-stability and are therefore triggered by the release
of neutrons at low values of the neutron density($n_n \lesssim 10^8 cm^{-3}$ . Actually, the
name {\it slow} derives exactly from this fact, as for low neutron densities most of the half-lives
of the unstable nuclei through which the nucleosynthesis chain passes are faster than neutron captures,
thus maintaining the path close to the stability valley. A second mechanism of neutron captures,
occurring at rapid rates (the ``r''-process) is necessary for producing about 50 \% of the abundances
for elements heavier than iron in supernovae, but its signatures have not been found so far in
presolar grains. It is now an accepted fact of stellar nucleosynthesis that the main part of
the s-process element distribution, from Sr to Pb (called ``the main component'') is produced
in AGB stars, through the recurrent operation of explosive He-shell burning and of TDU \cite{b99}

Most measurements of these trace elements in presolar materials referred to bulk samples of pristine
meteorites, where the presolar grain anomalous composition dominates. Heavy noble gases (krypton and
xenon) were measured by \cite{lew} and early interpreted in s-process models by \cite{ga1}. Strontium
and barium were studied by \cite{pod} and by \cite{ott} and \cite{pro}. Neodymium and samarium were
measured by \cite{zi1} and, together with dysprosium, by \cite{ric}. Later, single grain measurements
became available, for Sr, Zr and Nd \cite{nic1,nic2,nic3}. Then Ba was addressed by \cite{sav}.
Figure 4 shows examples of the expected He-shell composition, deduced from the measured data
through three-isotope plots, as compared to nucleosynthesis model calculations performed by \cite{gal98,ga2}, using variable efficiencies in the neutron production (different open symbols).
The elements Nd and Sm are considered here; the agreement is quite good and confirms beyond any doubt
the AGB origin of these grains. Notice that the models required correspond to AGB stars with a
metallicity higher than for an average model yielding the best fit to the solar main s-process
component \cite{gal98}. As the metallicity gradually increases with the chemical evolution of the
Galaxy, this suggests an origin of the grains in circumstellar envelopes of AGB stars born in a part of the Galaxy that was still relatively young when the Sun formed, i.e. belonging to the the Galactic disk.

Notice also that recent updates of the s-process \cite{mai11,mai12} suggested a significant expansion of the reservoir where the neutron source $^{13}$C($\alpha$,n)$^{16}$O operates. However, as these new suggestions
also indicate a higher average metallicity for the typical star fitting the solar s-process composition,
the number of neutrons per iron seed required to fit either the solar composition or the grains remains unchanged and we do not expect any significant change for model predictions like those displayed in Figure
4, at least for nuclei of A $>$ 90 (i.e. the main component). This guess is however now the object of a direct verification.

\section{Conclusions}
In this note we have illustrated, with the help of a few examples, the important role played by presolar grain measurements in constraining the models for AGB nucleosynthesis. While the level of agreement achieved in comparing iteratively model predictions and measured data is in general good, and guarantees that more 90\%
of presolar grains so far identified do come from AGB stars, there remain puzzling sources of concern,
especially for nitrogen. The large spread in the N isotopic ratio displayed by SiC grains, even by the mainstream ones that for the rest show unambiguously their AGB origin, is not reproduced by the models. It actually shows evidence of $^{15}$N enrichment, something that common nucleosynthesis knowledge attributes only to explosive phases, certainly not to red giants. The exploration of binary evolution and nucleosynthesis paths possibly leading to interactive, common-envelope phases that can be observationally misinterpreted as single stars is underway, in the hope that the grains with an anomalous composition measured can in fact come from such a scenario. The recent observation of peculiar C-rich red giants displaying the same problem for nitrogen isotopes, and the fact that some such stars are almost certainly binaries, encourages this line of research.

\begin{theacknowledgments}
O.T. and S.P. acknowledge support from the ``Istituto Nazionale di Fisica Nucleare'' (INFN).
E.M. acknowledges support from the ``Istituto Nazionale di Astrofisica'' (INAF)
The authors thank an anonymous referee for very helpful suggestions.
\end{theacknowledgments}



\bibliographystyle{aipproc}   



\end{document}